# Micropillar compression of single crystal tungsten carbide, Part 1: Temperature and orientation dependence of deformation behaviour

Helen Jones[1], Vivian Tong[1,*], Rajaprakash Ramachandramoorthy[2,#], Ken Mingard[1], Johann Michler[2], Mark Gee[1]

1. National Physical Laboratory, Hampton Road, Teddington, Middlesex TW11 0LW, United Kingdom

2. Empa – The Swiss Federal Laboratories for Materials Science and Technology, Feuerwerkerstrasse 39, 3603 Thun, Switzerland

## Abstract

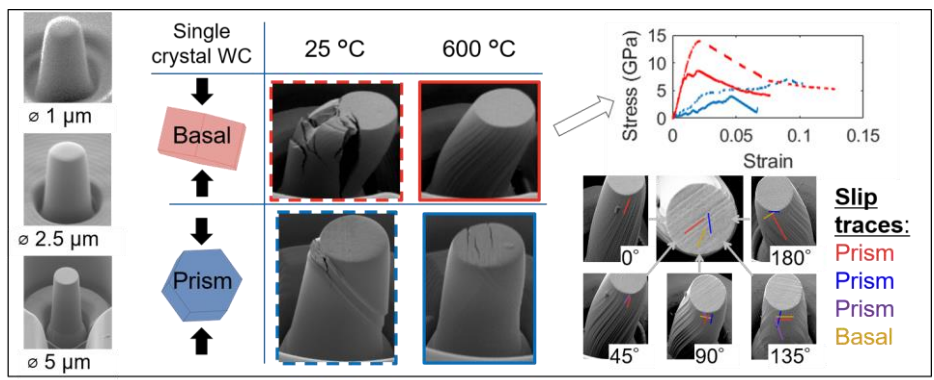

Tungsten carbide cobalt hardmetals are commonly used as cutting tools subject to high operation temperature and pressures, where the mechanical performance of the tungsten carbide phase affects the wear and lifetime of the material. In this study, the mechanical behaviour of the isolated tungsten carbide (WC) phase was investigated using single crystal micropillar compression. Micropillars in two crystal orientations, 1-5 µm in diameter, were fabricated using focused ion beam (FIB) machining and subsequently compressed between room temperature and 600 °C. The activated plastic deformation mechanisms were strongly anisotropic and weakly temperature dependent. The flow stresses of basal-oriented pillars were about three times higher than the prismatic pillars, and pillars of both orientations soften slightly with increasing temperature. The basal pillars tended to deform by either unstable cracking or unstable yield, whereas the prismatic pillars deformed by slip-mediated cracking. However, the active deformation mechanisms were also sensitive to pillar size and shape. Slip trace analysis of the deformed pillars showed that $\{10\bar{1}0\}$ prismatic planes were the dominant slip plane in WC. Basal slip was also activated as a secondary slip system at high temperatures.

---

[*] Corresponding author: vivian.tong@npl.co.uk

[#] Current affiliation: Max-Plank-Institut für Eisenforschung GmbH, Max-Planck-Strasse 1, 40237 Düsseldorf, Germany

# Highlights

- Single crystal WC micropillar compression from room temperature to 600 °C
- Basal directions 3 × stronger than prismatic directions in compression
- Micropillar shape artefacts affect cracking versus yielding behaviour
- Primary slip planes are $\{10\bar{1}0\}$ from room temperature to 600 °C
- Secondary slip on $(0001)$ plane activated at 600 °C

# 1 Introduction

## 1.1 Motivation

The high compressive strength and wear resistance of hardmetal components, used for machining and drilling applications, is derived from the hexagonal WC phase, which forms the principal component of the hardmetal when sintered, with typically 3-15 wt. % of a ductile binder phase such as Co or Ni [1]. Material simulations can be used to understand the behaviour of hardmetal composites under complex operating conditions and enable appropriate hardmetal grades to be selected or developed for each application. The majority of deformation models so far have focused on plastic deformation of the binder phase, where WC is approximated as a brittle elastic phase [2–10]. A recent *in situ* TEM study of observed that plastic deformation in hardmetals is accomodated by both WC and Co phases, even though final failure occurs within the binder phase [11,12]. However, the crystallographic nature of the lattice defect types contributing to plastic slip were not characterised in the TEM study.

WC grains soften significantly at high temperatures, showing a two- to six-fold decrease in indentation hardness between room temperature and 700 °C [13,14]. Although most of the deformation is accommodated by plastic flow of the binder phase ('cobalt drift'), grain boundary sliding, and interfacial cracking [9,13–21], microstructural examinations of deformed hardmetals have shown that the WC grains accommodate some plastic strain at both room and high temperatures [16,18]. An indentation study also showed different high-temperature softening behaviour between WC crystals and hardmetals [13].

Therefore, temperature- and strain rate-dependent mechanical properties of the isolated WC phase (as well as the binder and interfacial properties) are needed to accurately model the operational performance of hardmetal components, which experience temperatures of up to 1000 °C [22,23].

The temperature-dependent mechanical properties of isolated WC crystals, and the corresponding crystallographic slip and cracking mechanisms, are not well known. A wide range of room temperature slip systems and dislocation types in WC have been reported in published literature [13–15,17,20,21,24–33] and are listed in Supplementary Table 1, but only $\langle a \rangle$, $\langle c \rangle$, and $\langle c+a \rangle$ slip on $\{10\bar{1}0\}$ prismatic planes are consistently observed. In these reports, two methods have been used to characterise slip systems: residual dislocation analysis and slip trace analysis. The large scatter in observed slip systems is likely due to limitations of these methods when applied to WC.

Residual dislocation analysis infers the active slip systems from the dislocation populations of deformed materials using transmission electron microscopy (TEM) [17,29,34,35]. However, this method does not easily distinguish glissile dislocations, which accommodate plastic strain during deformation, from sessile dislocations, which are obstacles to glide and contribute to strain hardening and cracking. Sessile dislocations are found in as-sintered WC crystals, but can also nucleate during deformation from a variety of partial dislocation and stacking fault reactions [27,34].

Slip trace analysis measures the slip plane and direction from the orientation and step height of a slip band, ensuring that the measured slip system has been produced by material deformation. The dislocation Burgers vector is inferred from the slip direction.

However, inferring the dislocation type from the slip direction is ambiguous in WC, because multiple symmetric $\langle c + a \rangle$ dislocations can slip cooperatively on a $\{10\bar{1}0\}$ prismatic plane to allow slip in any in-plane direction, including the $\langle a \rangle$ and $\langle c \rangle$ directions. Therefore, $\langle a \rangle$ or $\langle c \rangle$ dislocation slip on prismatic planes cannot be distinguished from multiple $\langle c + a \rangle$ slip using slip trace analysis.

The activated slip systems at each temperature controls plasticity in WC. If only $\{10\bar{1}0\}$ prismatic slip is allowed, plastic deformation in WC is limited, as only four degrees of freedom are available, and the von Mises yield criterion for arbitrary deformation is not fulfilled [36,37]. However, if $\langle a \rangle$ dislocations could cross-slip onto secondary slip planes such as the basal plane, arbitrary plastic deformation would be possible, though there is no empirical observation of this at room temperature [25,38,39].

Our study aims to measure the anisotropic and temperature-dependent mechanical response of WC single crystals, and link this to the activated deformation mechanisms such as crystallographic slip and cracking. We address the ambiguity in conventional slip system characterisation methods by complementing slip trace analysis with electron backscatter diffraction (EBSD) mapping of deformed micropillar cross-sections in Part 2 of this work [40], and identify the primary slip systems by their characteristic lattice rotation axes.

## 1.2 Micromechanical testing

WC has an ordered hexagonal crystal structure and highly anisotropic mechanical properties [14,24,28,41–43]. Therefore, multiple tests on similar-oriented samples are needed to ensure that temperature and orientation dependence of mechanical properties are measured independently. Micromechanical testing is well-suited for this approach, as many small samples can be fabricated in a single crystal using a focused ion beam (FIB) [44–50]. Micromechanical testing can also suppress crack propagation in brittle materials [51,52], which is advantageous as WC is relatively brittle at room temperature.

In this study, three sizes of single crystal WC micropillars, fabricated in two crystallographic orientations which cover the range of mechanical anisotropy in WC, were tested at five temperatures between room temperature and 600 °C to investigate the temperature dependence of deformation and fracture behaviour. This builds on previous work using high temperature nanoindentation to measure the temperature and orientation dependence of hardness in WC grains [14].

Micropillar compression has some advantages over nanoindentation: firstly, the mechanical data from is easier to interpret, because of its nominally near-uniaxial loading stress and finite sample volume, whereas indentation produces a more complex stress state and plastic strain field; secondly, the deformed material is not embedded in a larger volume, so post-mortem characterisation methods can be used to link the mechanical response to the activated deformation mechanisms [49], such as the slip trace and cross-section EBSD analyses shown in Parts 1 and 2 of this work respectively.

However, there are significant experimental uncertainties in micropillar tests, especially at high temperatures. Micropillars fabricated by annular FIB milling suffer from edge rounding and side wall taper artefacts [53], so that the pillar stress state is not uniaxial. Pillar surfaces can also be damaged from ion beam irradiation [54–57]. In this work, a sharp focused beam and low ion current was used in the final fabrication steps to minimise beam damage, but there were still variations in pillar shape between nominally similar pillars and between different pillar sizes. During high temperature micropillar compression, attention was paid to temperature control, the sample and compression indenter tip were maintained at near identical temperatures to minimise thermal drift.

## 1.3 Microstructural analysis

Slip trace analysis of the deformed pillars was used to identify the active slip planes at room temperature, and additional slip planes activated at high temperatures. Slip directions are estimated by considering the geometrical alignment of candidate slip directions with the pillar loading axis (i.e. choose the slip system with maximum Schmid factor). However, as explained in Section 1.1, attempting to infer the dislocation slip direction(s) in WC from slip trace steps leads to ambiguous results, because $\langle a \rangle$ or $\langle c \rangle$ single slip cannot be distinguished from $\langle c + a \rangle$ multiple slip.

Part 2 of our study [40] will overcome this ambiguity by independently measuring the active primary slip system using electron backscatter diffraction (EBSD) lattice rotation mapping of deformed micropillar cross-sections. This method uses neither the slip traces nor the residual dislocation content, but instead identifies the primary slip system of the deformed pillar by their characteristic lattice rotation axes.

In addition to characterising WC slip, we also demonstrate a useful method to index crystallographic slip planes from curved slip traces on deformed cylindrical micropillars. Although slip plane indexing is more straightforward in cuboidal micropillars with flat side walls and linear slip traces, a cylindrical pillar geometry was chosen for this experiment to minimise stress concentrations near the corners of cuboidal samples during deformation [58], which is especially important when testing brittle materials such as WC.

# 2 Experimental method

## 2.1 Material preparation

Large WC crystals were grown and collected from the side walls of a hardmetal sintering furnace. Two single crystals, approximately 2 mm in width, were polished on triangular $(0001)$ basal and rectangular $\{10\bar{1}0\}$ prismatic WC facets respectively.

The crystals were embedded in resin with basal and prismatic facets flat on the top face, then ground from underneath to create parallel-sided samples. The top faces were mechanically polished using 9 µm, 3 µm and 1 µm diamond pastes followed by 30 nm colloidal silica. The samples were removed from the resin and mounted on a pin stub using Omegabond600 thermally conductive high temperature cement, which was then covered in a layer of high temperature electrically conductive paste (Pyro-Duct 597-A).

## 2.2 Pillar fabrication

Three arrays of twenty pillars, in the two WC crystals, with target top diameters of 1 µm, 2.5 µm and 5 µm and nominal aspect ratio of 2.5, were fabricated in two crystals orientations, using a Carl Zeiss Auriga-60 focused ion beam scanning electron microscope (FIB-SEM) and FIBICS nano-patterning system.

Annular milling with 30 kV Ga+ ions and progressively lower ion currents (4 nA, 600 pA and 240 pA) were used to minimise the side taper and ion damage to the pillar side walls. Rectangular viewing trenches were milled before annular milling to enable accurate pillar height measurement.

A 5 µm pillar array, and one example pillar of each size, are shown in Figure 1. The measured pillar top diameter varied from the target values by ±0.2 µm. The average side wall taper angles were between 5.9° and 2.3° and decreased with increasing pillar size. The pillar aspect ratios were between 1.8 and 2.3 and did not vary with pillar size. Pillar height measurement uncertainty, caused by tapering at the base of the pillar and milling artefacts from multiple passes with different FIB currents, contribute to scaling errors in engineering strain, especially in the 5 µm pillars where the base was not flat and the bottom of the pillar was less visible.

The crystal orientations of the polished surfaces were measured using EBSD and shown by hexagonal unit cells in Figure 1. The surface normal vectors of the the prismatic-oriented and basal-oriented crystals were [-1 20 -19 -9] (3° from [01-10]) and [3 5 -8 -51] (11° from [0001]) respectively. This corresponds to the loading direction of the fabricated micropillars.

The misalignment angles between the crystal surface and SEM stage axes were estimated by tilting the SEM stage to 70°, then measuring the apparent image rotation from applying 180° in-plane stage rotation. The component of surface misalignment perpendicular to the SEM tilt axis was then calculated using the method from [59]. Misalignment angles of 2.3° and 2.2° were measured for the prismatic-oriented and basal-oriented crystals respectively, resulting in an orientation measurement uncertainty of 6-7° for both crystals.

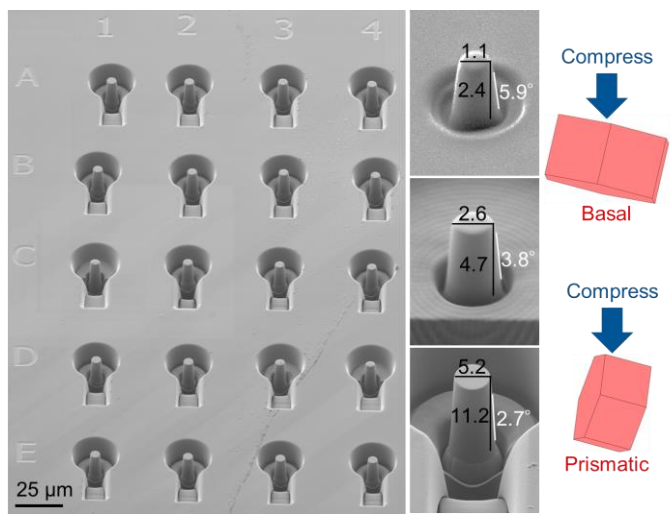

*Figure 1: Secondary electron image of an array of 5 μm diameter pillars before compression testing, with examples of individual 1 μm, 2.5 μm and 5 μm diameter pillars before deformation, showing average top diameter, height and taper measurements. The values in black have units of micrometres. The sample was tilted 54° to view the full height of the pillars and the image tilt-corrected by 36° to measure the height of the pillars.*

## 2.3  High temperature micropillar compression

An Alemnis *in situ* high temperature nanoindenter fitted inside a Carl Zeiss DSM962 tungsten filament SEM was used to perform the pillar compression tests. The nanoindenter had a maximum load of 1 N and maximum operating temperature of 800 °C.

The nanoindenter was tilted to 22° on the stage inside the microscope so that the pillars could be viewed during testing and the diamond flat-punch indenter tip could be positioned accurately over each pillar. Pillars were compressed using a diamond flat-punch indenter with a 10 μm diameter and 60° conical angle at the sides.

The Alemnis nanoindenter has a PID based feedback loop that allowed all the tests reported in this study to be conducted under true displacement control.  The tests were run for as long as possible and stopped manually just before catastrophic destruction of the pillars, e.g. where significant softening or large load drops were observed. Small misalignments between the indenter flat punch and the pillar top surface led to non-uniform loading of some of the pillars, which produced an initial toe region in the stress-strain curves.

Engineering stress and strain were calculated from the load and displacement signals through division by the micropillar top cross-sectional areas and pillar heights respectively. The data was also corrected for elastic sink-in of the pillar into the substrate using Sneddon's correction [60] and the indenter frame compliance specific to the high temperature version of the nanomechanical tester. The engineering strain rate measured after applying compliance corrections was 0.003 s$^{-1}$ for all pillars.

The high temperature tests were conducted using a heater and thermocouple combination attached to the sample and the diamond tip. Pillar compression experiments were conducted at five nominal temperatures: room temperature (approximately 25°C), 150°C, 300°C, 450°C and 600 °C. The temperatures of both the tip and sample were controlled using a PID feedback loop, based on the thermocouple readings. A temperature gradient was present between the sample surface and thermocouple, which was attached to the molybdenum sample stub, about 2-3 mm away from the sample surface. Therefore, in order to avoid thermal drift in the displacement due to temperature mismatch between the tip and the sample beyond the thermal stabilization period, careful temperature matching calibrations were conducted at each testing temperature for both samples. This calibration involved conducting a load-controlled flat punch indentation on the bulk sample surface with a drift hold segment. The tip and sample temperatures were then fine-tuned to match, based on achieving a low thermal drift of ~0.03 nm/s during the drift hold segment. Details of the instrumentation are reported in [61].

## 2.4 SEM slip trace analysis

Activated slip planes in the 5 μm pillar arrays were measured using slip trace analysis. Since crystallographic slip traces are planar sections through a cylindrical pillar, slip traces on the pillar top face are straight lines, and slip traces on the pillar side wall vary sinusoidally as the sample is rotated through 360°. The 2-3° side wall taper in 5 μm pillars (Figure 1), the ~2° sample surface misalignments, and the 6-7° surface orientation uncertainty (detailed in Section 2.2) were not considered in this analysis.

Slip traces on the pillar top faces were imaged at 0° stage tilt, and pillar side walls were imaged at 45° tilt from eight viewing directions, covering 360° in 45° increments. The crystallographic orientation of the sample was measured using EBSD at 70° stage tilt. The out-of-plane misalignment between the sample surface and SEM stage was measured during EBSD measurement using the 180° in-plane stage rotation method from [59].

Slip trace angles were measured from the pillar side walls by drawing tangent lines to curved slip traces at the front of the pillar, and measured on tilt-corrected micrographs in ImageJ [62]. Theoretical plane trace angles on the pillar top face and eight side wall viewing angles were computed in MTEX [63] and matched to the measured trace angles to identify the crystallographic slip planes. Examples of slip trace analysis on prismatic and basal pillars will be shown in Figure 6 and Figure 7 respectively.

# 3 Results and analysis

## 3.1 Mechanical data

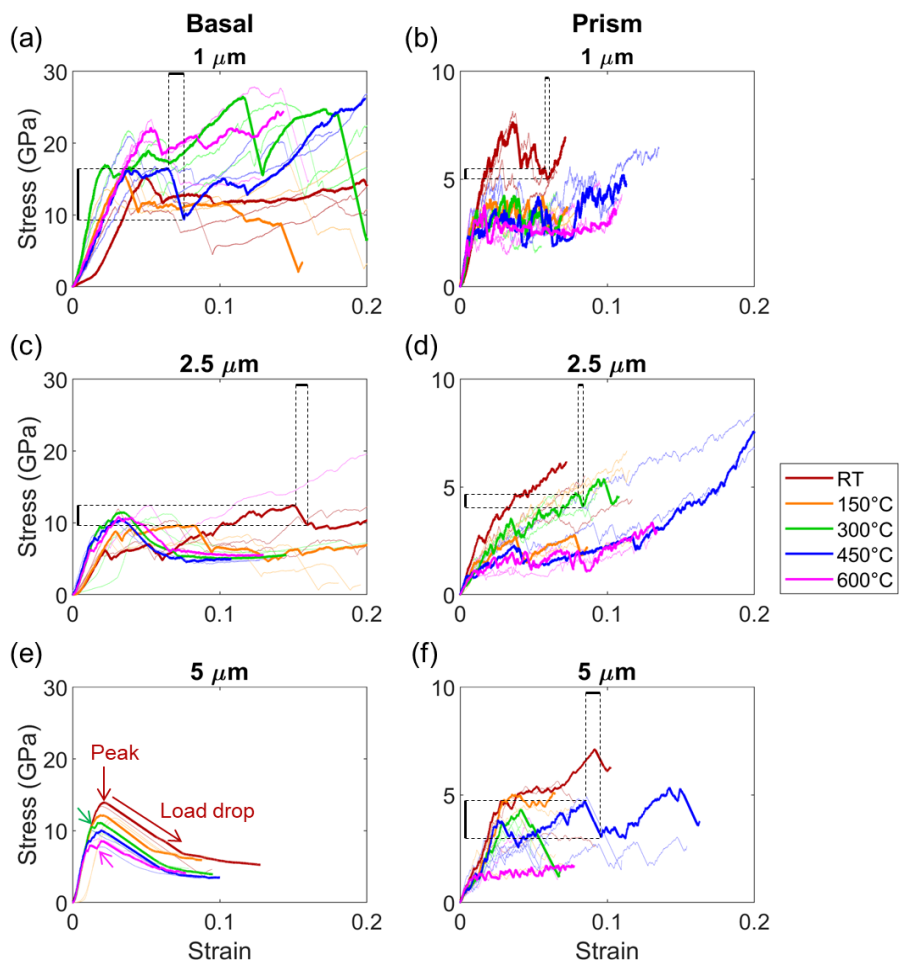

*Figure 2: Loading portion of stress-strain curves of all pillars, grouped by pillar size (rows), loading orientation (columns), and test temperature (plot colours). One line per pillar type is plotted in thicker lines to help visualization, and nominally identical repeats are drawn as thin lines. See text for details of annotation marks. Higher resolution stress-strain plots for all pillars, grouped by orientation and test temperature, are provided separately in Supplementary Information (B).*

Engineering stress-strain plots of the loading portions for all pillars are shown in Figure 2. Load-time curves, plotted in Supplementary Figure 1, show that the stress drops/strain jumps in Figure 2 are instantaneous load drops during the mechanical test, as expected from a displacement-controlled test.

The stress at first load drop for each pillar, corresponding to the first plastic deformation or crack initiation event, are plotted in Supplementary Figure 2. Both the plastic flow stresses and the first load drop stresses were about three times higher for basal than prismatic oriented pillars. Both orientations showed a 'smaller is stronger' size effect when comparing stresses in the 1 μm and 2.5 μm pillars, but negligible difference between the 2.5 μm and 5 μm pillars.

### 3.1.1 Basal oriented pillars

Basal-oriented pillar deformation could be classified as either stochastic or deterministic deformation. Deterministic deformation is characterised by smooth, reproducible stress-strain curves typically seen in macro-scale mechanical testing of ductile materials. Stochastic deformation is characterised by load drops due to intermittent plastic slip or crack propagation, and elastic deformation in between, and a wide scatter in flow stresses between nominally similar pillars. Stochastic plastic slip can occur in micropillars because few dislocation sources are available in a small sample volume, which limits the number of glissile dislocations available for slip.

Stochastic deformation was observed in smaller pillars deformed at lower temperatures: 1 µm pillars deformed at all temperatures, and 2.5 µm pillars deformed at 150 °C and at room temperature. The stress-strain curves contained intermittent load drops and wide scatter in flow stress between similar pillars. Two characteristic groups of load drops are visible in the pillars: large drops (2 - 10 GPa) likely related to cracking, and smaller drops (< 0.5 GPa) likely related to dislocation slip. It will be shown in Figure 4 that large load drops are correlated with severe cracking in the deformed pillars. No significant temperature dependence was observed in stochastically deformed basal pillars.

Deterministic deformation was observed in larger pillars deformed at higher temperatures: 5 µm pillars at all temperatures, and most of the 2.5 µm pillars deformed at 300, 450 and 600 °C. The stress-strain curves were repeatable between pillars: an initial peak at around 2 % compressive strain, followed by a single large load drop to 7 % strain (Supplementary Figure 1f), then a nearly flat region after a 'knee' in the stress-strain curve. An example peak and load drop is marked for the RT basal pillar by the brown arrows. Unlike load drops in the stochastically deformed samples, the large load drop in these pillars was not followed by a corresponding load increase. The green and pink arrow annotations in Figure 2e show small horizontal steps, most likely stochastic dislocation nucleation events, in some pillars deformed between 300 and 600 °C, just before the main peak at 2 % strain. Flow softening with increasing temperature was observed in the 5 µm pillars. In the 2.5 µm pillars, flow stresses did not vary much between temperatures of 300 °C and 600 °C.

### 3.1.2 Prismatic oriented pillars

All prismatic-oriented pillars deformed by stochastic plastic flow, with small fluctuating load drops around an overall increasing load. Increasing the temperature from RT to 600 °C led to a two or three-fold decrease in flow stresses for all pillar sizes. However, the first load drop stresses (Supplementary Figure 2b) of prismatic pillars of all sizes showed little temperature dependence between RT and 600 °C.

As the pillar size increased, the load drops increased in both stress amplitude and strain accommodated per drop. This can be seen in the pairs of horizontal and vertical dashed black lines in Figure 2, which mark the start and end positions of a 'typical' stress drop. The dashed lines are intended only as a qualitative visual guide, and a wide range of stress drops can be readily observed. The maximum stress drops measured in each pillar are shown later in Figure 4 and Figure 5.

## 3.2 Stress-strain gradient plots

Figure 3 shows the change in stress per unit strain for all pillars. This was used to visualise trends between groups of pillars that were not obvious from the stress-strain plots, since stochastic load drops, which cause a large scatter in flow stress (Figure 2), are reduced to sharp spikes in these plots. Similar plots have been used in macro-scale mechanical tests to compare the strain-hardening behaviour of polycrystals, such as Figure 1(b) in Reference [64], where inflection points in the stress-strain gradients indicate transitions between slip- versus twinning-dominated plastic deformation in Zr. In the analogous plots in Figure 3, groups of different pillar deformation modes can be visualised. Details of how these plots were calculated are described in the caption of Supplementary Figure 3.

The scatter range of basal pillars widens with decreasing pillar size as the deformation tends towards stochastic flow. The strain hardening plots in Figure 3 show that basal pillars of all sizes deformed in three distinct regimes: (1) a large positive stress-strain gradient (400 to 800 GPa) below 2 % strain corresponding to 'elastic loading' before the first load drop, (2) a negative stress-strain gradient (−50 to −200 GPa) between 2 and 6% strain, and (3) near zero gradient (−50 to +100 GPa) beyond 6 % strain. The average values in these three regimes vary slightly between pillar sizes, but are independent of temperature.

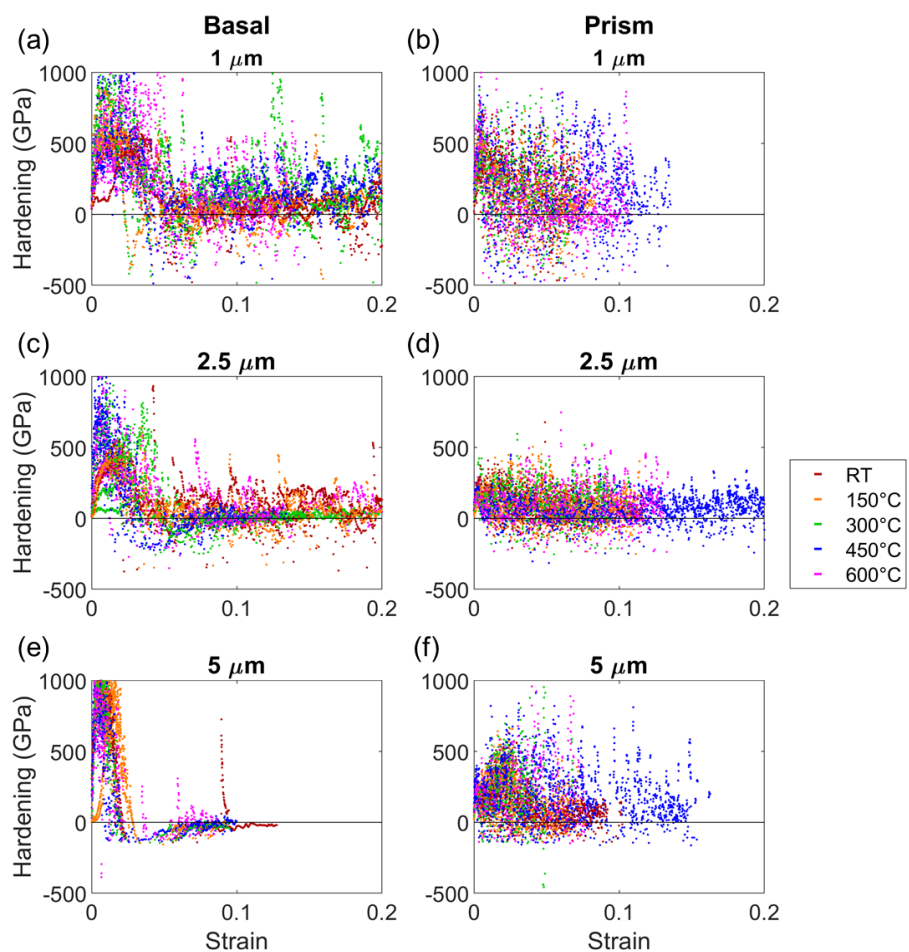

*Figure 3: Stress-strain gradient plots of the micropillars, measured from local gradients of stress-strain curves. Only the loading portions of the stress-strain curves are used. Data points from similar pillars are plotted together and not distinguished.*

The average stress-strain gradient of prismatic pillars was near-constant (100 - 200 GPa) after the initial 'elastic loading' region below 2 % strain, and independent of deformation temperature or pillar size. The scatter range, corresponding to the degree of stochasticity, was large for 1 µm pillars and decreased for larger pillars. The stress-strain gradients at load drops (negative spikes) in 2.5µm and 5 µm pillars were smaller (about −100 GPa), in contrast to the 1 µm pillars which had many load drops with stress-strain gradients beyond −300 GPa. This suggests that the origin of load drops in 1 µm pillars could be different to the 2.5µm and 5 µm pillars and will be explored further in Section 3.3.

## 3.3   Cracking behaviour

SEM imaging of the deformed pillars was used to characterize the degree of cracking. Representative micrographs of the pillars, imaged from two viewing angles, are shown in Figure 4 (basal pillars) and Figure 5 (prismatic pillars). Where different types of deformation patterning were observed in nominally similar pillars (e.g. 1 µm / 150 °C basal pillars in Figure 4), representative SEM micrographs are shown for each type. Larger versions Figure 4 and Figure 5 are provided as a supplementary document file to visualise SEM image details.

The stochasticity of stress-strain data and size of the largest stress drop (Figure 2) are tabulated alongside the SEM images to show correlations between cracking severity, flow stochasticity, and stress drop size. In basal oriented pillars, the first large stress drop near the yield point was ignored as this was observed in all pillars, even those that did not contain any cracks.

| Temp. / °C | Max stress drop / GPa | Stochastic? | 1 μm | Max stress drop / GPa | Stochastic? | 2.5 μm | Max stress drop / GPa | Stochastic? | 5 μm |
|---|---|---|---|---|---|---|---|---|---|
| 25 | 8.5<br>6.4<br>5.5 | 1<br>1<br>1 | | 6<br>2.6<br>0.4 | 1<br>1<br>1 | | 0.8<br>0.05 | 0<br>0 | |
| 150 | 0.4<br>5.2<br>6.6 | 1<br>1<br>1 | | 5.85<br>2.7<br>3.45 | 1<br>1<br>1 | | 0.12<br>0.02<br>0.15 | 0<br>0<br>0 | |
| 300 | 17.5<br>10<br>10 | 1<br>1<br>1 | | 0.25<br>0.5<br>0.5 | 0<br>0<br>0 | | 0<br>0<br>0.5 | 0<br>0<br>0 | |
| 450 | 2.3<br>3.4<br>5.5 | 1<br>1<br>1 | | 0.5<br>0.4<br>0.2 | 0<br>0<br>0 | | 0<br>0 | 0<br>0 | |
| 600 | 3<br>2.5<br>12 | 1<br>1<br>1 | | 0.5<br>5.9 | 0<br>1 | | 0.3<br>0.4<br>0.4 | 0<br>0<br>0 | |

*Figure 4: Relationship between cracking and stochastic stress-strain behaviour in basal oriented micropillars. The unit cells show the crystal orientation on the pillar front wall, and have been vertically rescaled to match the angles at 45° sample tilt used for imaging these pillars. The background colours of the 'Max stress drop' cells are formatted to match the stress drop size: 0 GPa = green, 3 GPa = yellow, ≥ 7 GPa = red. A larger version of this table is provided separately in Supplementary Information (A).*

The basal pillars showed two types of characteristic cracking modes: unstable crack propagation and catastrophic fracture in small pillars deforming at low temperatures, and ductile plastic deformation and stable crack growth in large pillars at high temperatures.

Section 3.1.1 described how basal pillars can be categorised by either stochastic or deterministic stress-strain behaviour. Figure 4 shows that stochastically deformed pillars all cracked severely along $\{10\bar{1}0\}$ planes (red dashed lines), and the largest stress drops were typically > 5 GPa. In contrast, deterministically deformed pillars showed minimal cracking, but instead severe pillar bending, resulting in curved slip traces. The pillar shapes qualitatively resemble the single crystal micropillars in [65] which had an initial unstable orientation, and deformed by plastic buckling. (An unstable crystal orientation is one that tends to rotate away from its initial orientation during deformation [66].) The cracks which formed in these pillars are in an 'X' shape in the tensile part of the bent pillar wall. The crack traces appear to lie along $\langle c + a \rangle$ directions (blue and green dashed lines), although this is uncertain due to the severe bending in these pillars.

The crack plane could be uniquely identified where surface traces were visible on both the top face and side wall, such as the prismatic crack plane marked by red dashed lines in Figure 4 (1 µm pillars, 150 °C). The crack plane could be partially identified where only one surface trace is visible, such as the 'X'-shaped cracks marked by blue and green dashed lines in Figure 4 (2.5 µm pillars, 300 °C); in these cases, the crack planes must contain the $\langle c + a \rangle$ surface trace directions.

| Temp. /°C | Max stress drop / GPa | Stochastic? | 1 µm | | Max stress drop / GPa | Stochastic? | 2.5 µm | | Max stress drop / GPa | Stochastic? | 5 µm | |
|---|---|---|---|---|---|---|---|---|---|---|---|---|
| 25 | 1.9 | 1 | | | 0.2 | 1 | | | 0.2 | 1 | | |
| | 2.1 | 1 | | | 0.45 | 1 | | | 0.7 | 1 | | |
| | 1.3 | 1 | | | 0.2 | 1 | | | 0.5 | 1 | | |
| 150 | 0.9 | 1 | | | 0.6 | 1 | | | 0.4 | 1 | | |
| | 1.3 | 1 | | | 0.45 | 1 | | | 2.5 | 1 | | |
| | 0.9 | 1 | | | 0.3 | 1 | | | 0.16 | 1 | | |
| | | | | | 0.8 | 1 | | | 0.25 | 1 | | |
| 300 | 1.2 | 1 | | | 1 | 1 | | | 2.5 | 1 | | |
| | 1.1 | 1 | | | 0.5 | 1 | | | 0.5 | 1 | | |
| | 0.8 | 1 | | | 0.8 | 1 | | | 0.6 | 1 | | |
| 450 | 1.6 | 1 | | | 1.3 | 1 | | | 1.6 | 1 | | |
| | 1.4 | 1 | | | 0.4 | 1 | | | 1.7 | 1 | | |
| | 1.2 | 1 | | | 0.7 | 1 | | | 1.1 | 1 | | |
| | | | | | | | | | 1.3 | 1 | | |
| 600 | 1.7 | 1 | | | 0.8 | 1 | | | 2.9 | 1 | | |
| | 0.6 | 1 | | | 0.4 | 1 | | | 0.4 | 1 | | |
| | 0.5 | 1 | | | 0.94 | 1 | | | 1.5 | 1 | | |
| | | | | | | | | | 0.6 | 1 | | |
| | | | | | | | | | 3.4 | 1 | | |

*Figure 5: Relationship between cracking and stochastic stress-strain behaviour in prismatic oriented micropillars. The unit cells show the crystal orientation on the pillar front wall and have been vertically rescaled to match the 45° sample tilt used for imaging these pillars. The background colours of the 'Max stress drop' cells are formatted to match the stress drop size: 0 GPa = green, 3 GPa = yellow, ≥ 7 GPa = red. A larger version of this table is provided separately in Supplementary Information (A).*

In the prismatic pillars, cracking severity increases with pillar size. No temperature dependence was observed in the types of cracks that formed, nor the cracking severity.

All the 1 µm pillars deformed by planar shear on a single slip plane containing an $\langle a \rangle$ direction (blue dashed line in unit cell schematic and on images), and some pillars had a small crack lying in the slip plane. In the 5 µm pillars, double slip was activated in most of the pillars, with slip traces along two $\langle a \rangle$ directions, and cracks propagated along the intersection of the two activated slip planes (orange dashed lines). Some pillars contained several cracks across the pillar width, but none failed catastrophically. This suggests that cracking in the prismatic pillars was stable and slip-mediated. The behaviour of the 2.5 µm pillars was somewhere in between, showing both single slip and double slip in different pillars.

Double slip activation increased the average flow strength of the pillars: out of three 5 µm prismatic pillars deformed at room temperature, double slip and cracking occurred in two pillars, and single slip with minimal cracking in one pillar. The corresponding stress-strain curves (Figure 2f, brown lines) showed that the double-slipped pillars had an average plastic flow stress of 5 GPa, whereas the single-slipped pillar had a lower average plastic flow stress, around 3 GPa.

## 3.4 Slip trace analysis

Activated slip planes in the 5 µm pillar arrays were measured using slip trace analysis at each temperature. Figure 6 and Figure 7 show slip trace measurements on two 5 µm pillars compressed at 600 °C, in prismatic and basal orientations respectively. The SEM viewing directions are shown schematically as keyhole-shaped drawings of the pillar and viewing trench, and slip trace angle sign conventions are defined in the pillar side wall and top face schematics in Figure 6a and Figure 7a. High resolution copies of the SEM images in Figure 6 and Figure 7 are provided separately in Supplementary Information (B).

### 3.4.1 Prismatic oriented micropillars

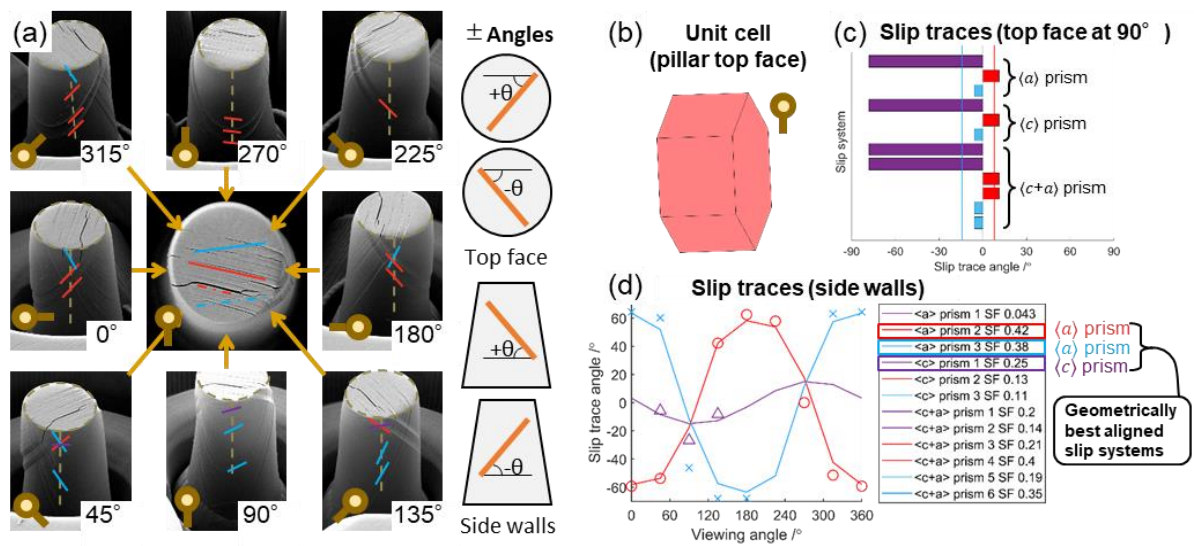

*Figure 6: Slip trace analysis of 5 µm prismatic micropillar compressed at 600 °C. (a) SEM images of the pillar top face and side walls at different viewing angle, showing three sets of crystallographic slip traces (solid lines) and cracks (dashed lines), marked in red, blue and purple. (b) Unit cell showing crystal orientation from EBSD of the undeformed sample. (c) Closest matching slip traces on the pillar top face: bars show expected slip trace angles from EBSD orientation; lines are experimentally measured from (a). (d) Closest matching slip traces on the pillar side walls: lines show expected slip trace angles from EBSD orientation; scattered points show experimental measurements for red, blue and purple slip traces in (a).*

The red, blue and purple lines in Figure 6a show slip traces (solid lines) and crack paths (dashed lines) corresponding to the three sets of slip planes measured from the images. Figure 6c compares measured slip trace angles (vertical lines) to the closest matching theoretical plane traces (bars) on the pillar top face. Figure 6d compares the measured (markers) with closest matching slip traces (sinusoid lines) expected for the pillar side walls. The three sets of slip traces can be indexed as $(\bar{1}100)$ (red), $(\bar{1}010)$ (blue), and $(01\bar{1}0)$ (purple) planes. All reported WC slip planes (listed in Supplementary Table 1) were considered during slip trace analysis, but only the indexed slip planes are plotted in Figure 6.

A few degrees of difference between the measured and expected slip traces is reasonable. Image distortions between slip trace imaging and EBSD orientation measurement, misalignment between the sample surface and SEM stage, lattice rotations from pillar deformation, pillar side wall taper, and uncertainty in identifying the front of the pillar all contribute to measurement uncertainty. Even with a ~2° misalignment between the sample surface and SEM axes, the slip traces could still be confidently indexed because consistent solutions were found from nine measurements (an arbitrarily oriented plane has two degrees of freedom), and because the crystallographic slip planes in WC are well-separated.

The geometrically best-aligned slip system for each slip plane was determined by calculating the slip direction with maximum Schmid factor, out of the possible slip directions listed in Supplementary Table 1. The legend in Figure 6d shows that the geometrically best-aligned slip systems were $[11\bar{2}0](\bar{1}100)$ (red slip traces, Schmid factor 0.42), $[1\bar{2}10](\bar{1}010)$ (blue slip traces, Schmid factor 0.38), and $[0001](01\bar{1}0)$ (purple slip traces, Schmid factor 0.25).

### 3.4.2 Basal oriented micropillars

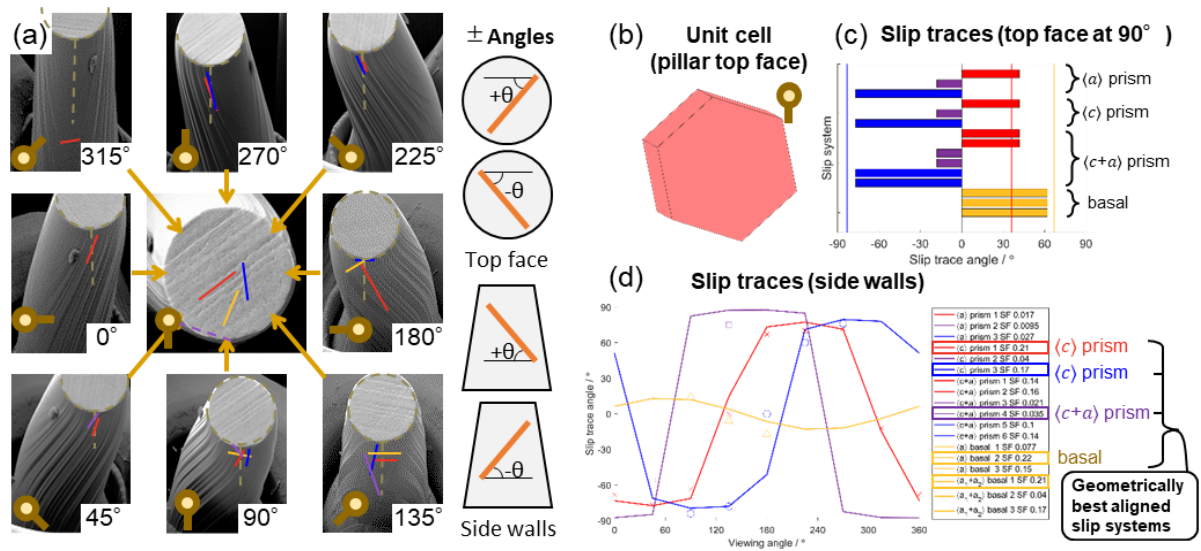

*Figure 7: Slip trace analysis of 5 μm basal micropillar compressed at 600 °C. (a) SEM images of the pillar top face and side walls at different viewing angles, showing three sets of crystallographic slip traces (solid lines) marked in red, blue and yellow. The dashed purple line on the top edge of the pillar shows a theoretical trace expected from the pillar side wall slip traces, but not was not directly observed. (b) Unit cell showing crystal orientation from EBSD of the undeformed sample. (c) Closest matching slip traces on the pillar top face: bars show expected slip trace angles from EBSD orientation; lines are experimentally measured from (a). (d) Closest matching slip traces on the pillar side walls: lines show expected slip trace angles from EBSD orientation; scattered points show experimental measurements for the slip traces in (a).*

Figure 7 shows the same slip trace analysis method applied to a [0001] oriented 5 μm pillar compressed at 600 °C. The slip traces were twisted into 'S'-shapes about the pillar loading axis, and the top of the deformed pillar was rotated away from the base. In contrast, the $[10\bar{1}0]$ pillar in Figure 6 showed planar slip and minimal plastic rotation. Slip traces on the pillar side walls were measured near the top of the pillar to minimise the error contribution from lattice rotation during slip trace analysis (see Appendix A).

Measured slip traces in this pillar corresponded to $(01\bar{1}0)$ (red), $(\bar{1}010)$ (blue), $(\bar{1}100)$ (purple), and $(0001)$ (yellow) slip planes. Most of the slip traces were along $(01\bar{1}0)$ (red) and $(\bar{1}010)$ (blue) planes, so these were most likely the primary slip systems; the $(\bar{1}100)$ (purple) and $(0001)$ (yellow) slip planes are likely secondary slip systems. Schmid factor analysis showed that none of the candidate slip directions on the measured planes were particularly well-oriented for slip, with Schmid factors ≤ 0.22. The best aligned slip systems were $[0001](01\bar{1}0)$ (red, Schmid factor 0.21), $[0001](\bar{1}010)$ (blue, Schmid factor 0.17), $[\bar{1}\bar{1}23](\bar{1}100)$ (purple, Schmid factor < 0.1), and $[11\bar{2}0](0001)$ (yellow, Schmid factor 0.22).

### 3.4.3 Additional slip systems activated at high temperature

One pillar per temperature and orientation was analysed using the method in Section 3.4. Table 1 shows the slip planes activated in each pillar, where additional slip planes activated at high temperatures are marked in bold font. The basal pillar deformed at RT had cracked during testing, which tilted large portions of the pillar and increased the measurement uncertainty; two sets of secondary slip traces could not be assigned to a crystallographic plane. All sets of primary slip traces were assigned successfully.

$\{10\bar{1}0\}$ prismatic planes were the dominant slip planes in the basal-oriented pillars despite relatively poor geometric alignment, confirming that prismatic planes are the only dominant primary slip plane in WC between RT and 600 °C.

| Orientation | Room Temp. | 150 °C | 300 °C | 450 °C | 600 °C |
|---|---|---|---|---|---|
| Prismatic | $[11\bar{2}0](\bar{1}100)$ $[1\bar{2}10](\bar{1}010)$ | $[11\bar{2}0](\bar{1}100)$ $[1\bar{2}10](\bar{1}010)$ | $[11\bar{2}0](\bar{1}100)$ $[1\bar{2}10](\bar{1}010)$ | $[11\bar{2}0](\bar{1}100)$ $[1\bar{2}10](\bar{1}010)$ | $[11\bar{2}0](\bar{1}100)$ $[1\bar{2}10](\bar{1}010)$ **$[0001](01\bar{1}0)$** |
| Basal | $[0001](01\bar{1}0)$ $[0001](\bar{1}010)$ | $[0001](01\bar{1}0)$ $[0001](\bar{1}010)$ | $[0001](01\bar{1}0)$ $[0001](\bar{1}010)$ **$[\bar{1}\bar{1}23](\bar{1}100)$** | $[0001](01\bar{1}0)$ $[0001](\bar{1}010)$ **$[\bar{1}\bar{1}23](\bar{1}100)$** | $[0001](01\bar{1}0)$ $[0001](\bar{1}010)$ **$[\bar{1}\bar{1}23](\bar{1}100)$** **$[11\bar{2}0](0001)$** |

*Table 1: Slip planes activated during micropillar compression as a function of test temperature and orientation. The expected slip direction is calculated by maximising Schmid factor in the undeformed pillar orientation.*

## 4 Discussion

### 4.1 Effect of pillar shape: unstable cracking versus slip in basal pillars

Over the range of temperatures and pillar sizes tested, the deformation behaviour of basal pillars changed between deterministic and stochastic flow. The stress-strain curves in Figure 2, and stress drop data and SEM images in Figure 4 show that larger pillars deformed at higher temperatures favoured ductile plastic deformation with stable cracking, deterministic flow, and small or no load drops. Smaller pillars deformed at low temperatures favoured unstable crack propagation, stochastic flow, and large load drops.

The choice between ductile plastic deformation versus unstable cracking could depend on other factors as well as the nominal pillar size and loading temperature. Pillar-specific parameters such as pillar taper, edge rounding, and indenter tip alignment can also affect the local stress state, which changed the activated deformation modes.

For example, Figure 4 shows two nominally similar 2.5 µm basal pillars compressed at 600 °C, which deformed by different modes. One pillar had cracked catastrophically by stochastic flow (Figure 2c, thin pink line), whereas the other pillar deformed by plastic buckling and deterministic flow (Figure 2c, thick pink line). SEM images before deformation (Supplementary Figure 5) showed that the pillar that cracked catastrophically had anomalously rounded top edges from FIB milling, which would have affected both the FIB damaged layer thickness and the local stress state.

The effect of pillar shape parameters can also be seen in the proportional loading parts of the stress-strain curves in Figure 2c, which are divided between either stochastic or deterministic flow, independent of test temperature. All deterministic stress-strain curves in Figure 2c had a higher proportionality constant of about 600 GPa, whereas all stochastic curves had a lower proportionality constant of about 400 GPa. (For comparison, the Young's moduli of single crystal WC, calculated from elastic constants in Lee and Gilmore [43,67], are 820 GPa and 600 GPa in the basal and prismatic directions respectively.) Previous finite element model simulations [53] have showed that increasing the edge rounding in micropillars (1) decreases the measured proportionality constant, (2) amplifies the local compressive stress where the indenter contacts first at the pillar top centre, and (3) produces corresponding radial tensile stresses which work to open cracks on the pillar top face. This explains why, out of the two pillars tested under nominally similar conditions, the pillar with more edge rounding deformed by catastrophic cracking, and the other by ductile plastic deformation.

In Figure 4, catastrophic cracking is observed in small basal pillars, but suppressed with increasing pillar size, which is the opposite to the typical size effect observed in brittle ceramics, where cracking is suppressed and ductility increases below a threshold size [68,69]. The apparent size effect observed in the basal pillars is better explained as a pillar shape effect: rounded edges occupy a larger fraction of the cross-section in smaller pillars, and edge-rounding produces high axial compressive and radial tensile stresses in the top centre of the pillars, so that cracks opened outwards during pillar compression. Smaller pillars also had a larger taper angle, so more of the axial loading stress is accomodated by the top part of the pillar, where the cross-section area is smaller.

This shows that the basal pillars are not intrinsically brittle, even at room temperature. However, the ductile yield strength is much greater than the fracture strength at elevated temperatures, whereas at room temperature they are comparable in magnitude.

## 4.2 Origin of load drops and stochastic deformation events

*In situ* imaging during micromechanical tests was not possible due to limited resolution at high temperatures in the tungsten filament SEM, so the load drops and other features in the stress-strain data could not be directly correlated to individual pillar deformation events.

Even so, variation in deformation mechanisms could be linked to the mechanical behaviour through representative post-test SEM images and the mechanical data, as summarised in Figure 4 and Figure 5.

In the basal pillars, the smaller pillars underwent catastrophic fracture, but the larger pillars deformed by plastic yielding. Catastrophically fractured pillars tended to have large stress drops > 5 GPa, whereas plastically yielded pillars had relatively smooth stress-strain curves with stress drops of no more than 0.5 – 1 GPa. The 'maximum stress drop' value reported in Figure 4 excluded the first large load drop observed in the basal pillars deformed in a ductile manner (e.g. Supplementary Figure 1e), which was seen in pillars that did not crack at all, and unlike all other load drops, was not followed by a steep load increase. This load drop likely corresponds to unstable yield or plastic buckling, caused by the pillar softening as it deformed. Therefore, the deformed pillars were bent, and the slip traces appeared twisted (e.g. Figure 7, 5 µm pillars in Figure 4).

In the prismatic pillars, cracking was more severe in the 5 µm pillars than the 1 µm pillars (Figure 5). This can be linked to different types of load drops in the mechanical data: Figure 2a and e show that the load drops in the 5 µm prismatic pillars accommodate more compressive strain than the 1 µm prismatic pillars. The stress strain gradients of load drops in the 2.5 and 5 µm basal pillars (negative spikes in Figure 3c, e) are also shallower than the stress-strain gradients of load drops in 1 µm pillars (Figure 3a). However, the maximum load drop size (Figure 5) did not vary strongly with pillar size.

The SEM and mechanical data can be combined to infer that the load drops in 1 µm pillars were related to stochastic dislocation activity, and possibly the formation of stacking faults or twins, which have been observed to cause similar stress-strain 'ripples' during bending of a WC-Co hardmetal thin foil [12]. In contrast, the load drops in 5 µm pillars resulted in a combination of dislocation activity and crack propagation.

Larger prismatic pillars were more likely to activate double slip, probably due to more dislocation sources available in the larger pillar volume. Double-slipped pillars were also more likely to contain cracks along the intersection between the two slip planes. Since the crack ran along the intersection of two slip planes, the crack growth was stable, as the stress ahead of the crack tip could be dissipated through dislocation slip.

In summary, stochasticity in the stress-strain curves of basal pillars were related to unstable cracking, whereas stochasticity in the prismatic pillars were related to dislocation slip and slip-mediated stable crack growth. Plastic buckling with deterministic flow was observed in basal pillars as an alternative deformation mode to catastrophic cracking.

## 4.3 Comparison to existing studies

### 4.3.1 Activated slip systems

The slip trace analysis results in Section 3.4 confirm literature reports that slip in WC occurs predominantly on $\{10\bar{1}0\}$ planes. The slip traces of the deformed pillars in this study were qualitatively similar to the 2 µm diameter WC micropillars shown by Csanádi [49]. In this study, additional secondary slip planes were also activated at high temperatures: a third $\{10\bar{1}0\}$ slip plane in the prismatic pillars at 600 °C and the basal pillars at ≥ 300 °C, and $\{0001\}$ slip in the basal pillars at 600 °C.

Slip on only prismatic planes cannot accommodate compressive strain along [0001]. However, prismatic slip was activated for the basal-oriented pillars in this study, because the pillar loading axes were about 11° misoriented from [0001].

From the present data, we expect prismatic slip to be the dominant slip plane for loading directions ≥ 11° from [0001], but cannot extrapolate this to loading directions closer than 11° from [0001]. This is consistent with the basal-oriented, RT deformed, WC micropillars of Csanádi et al. [49], which showed no signs of plastic deformation before brittle failure.

Slip on the (0001) plane has not been observed in WC literature, although Greenwood and Loretto observed $\langle a \rangle$ edge dislocations in the basal plane [25]. Density functional theory simulations from Nabarro et al. [38,39] predicted a Peierls stress only twice that of $\langle a \rangle$ dislocations on $\{10\bar{1}0\}$, so considering their analysis alone, basal slip could be expected as a secondary slip system. Out of the reported WC slip directions and dislocation types (Supplementary Table 1), $\langle a \rangle$ dislocations have a Burgers vector in the basal plane and therefore can cross-slip onto the basal plane. This would provide the fifth degree of freedom required for arbitrary plastic deformation, as described in Jayaram et al. [36]. Therefore, basal slip activation at 600 °C, as observed in this study, could partly contribute to the abrupt ductility increase observed in hardmetals at 600 °C and above [18].

### 4.3.2 Temperature dependence

Literature studies [13,14] report that indentation hardness is higher on basal planes than prismatic planes, which is consistent with the flow stress anistropy in Figure 2. These studies also observed a two to three-fold reduction in hardness between RT and 600 °C [13,14]. This is consistent with the prismatic micropillar flow stresses (Figure 2b, d, f), but not the first load drop stresses (Supplementary Figure 2b), which were only weakly temperature dependent. This might be an artefact of the non-cylindrical micropillar shapes in this study, which strongly affects the local stress distribution during elastic loading, but less so after the onset of plastic deformation.

Increased cracking at nanoindentation temperatures ≥ 500 °C, observed by de Luca et al. [14], was also absent in this study. A significant difference between deformation conditions in nanoindentation and micropillar compression is that during nanoindentation, the majority of dislocations remain inside the sample volume, whereas in micropillar compression, glissile dislocations can be emitted at slip steps on the side walls. Therefore, the increased cracking during high temperature nanoindentation could be caused by an increase in dislocation interactions in the larger plastically deformed volume. This is consistent with the prismatic pillars in this study, where intersecting slip bands in double-slipped pillars led to slip-mediated crack propagation.

The anomalous increase in strength of 1 μm basal pillars as deformation temperature increases from 450 °C to 600 °C, shown in Supplementary Figure 2, is neither seen in literature reports, nor reproduced in prismatic pillars or larger basal pillars in this study. The reason for this is unknown from the present data and left as an open question for future investigations.

# 5 Conclusion

Micropillar compression has been used to measure the anisotropic and temperature-dependent deformation behaviour of tungsten carbide single crystals. The pillars were approximately three times stronger when compressed along near-basal orientations compared to near-prismatic orientations.

Smaller basal pillars at low temperatures deformed by unstable cracking, larger basal pillars at high temperatures deformed by unstable plastic deformation, and prismatic pillars deformed by planar slip and slip-mediated stable cracking.

Basal slip was observed as a secondary slip system active at 600 °C. Basal slip activation could be important for understanding the increased ductility of hardmetals above 600 °C, as it provides the fifth degree of freedom required to satisfy the Von Mises criterion and enable arbitrary plastic deformation in WC. $\langle a \rangle$ basal slip has been suggested as a possible slip system but had not previously been experimentally observed.

The dominant slip plane was $\{10\bar{1}0\}$ for temperatures between RT and 600 °C, and orientations with loading direction ≥ 11° from $[0001]$. The activated deformation modes were anisotropic and temperature dependent, and were also sensitive to pillar size and shape, notably side wall taper and edge rounding, which cause non-uniaxial pillar loading, stress gradients within each pillar, and variation in stress state between pillars.

# 6 Acknowledgements

The Department of BEIS (Department for Business, Energy and Industrial Strategy) is thanked for funding from the NMS (National Measurement System) program. Members of the EPMA hard materials group (European Powder Metallurgy Association) are thanked for supplying the material. Thank you to Manish Jain at Empa for helping with indentation. Dr. Ramachandramoorthy would like to acknowledge the funding from the EMPAPOSTDOCS-II programme that has received funding from the European Union's Horizon 2020 research and innovation programme under the Marie Skłodowska-Curie grant agreement number 754364. Francois De Luca is also thanked for internal review of the manuscript before submission.

# 7 Author Contributions

HJ prepared the micropillar samples; HJ and RR performed the compression tests under guidance from JM; HJ and VT performed the SEM imaging. VT analysed the mechanical data using MATLAB scripts written by HJ and VT. VT performed the slip trace analysis. KM supervised the work throughout. VT drafted the manuscript with input from KM, HJ, RR and MG. MG and JM inspired the project by initiating a collaboration between Empa and NPL.

# 8 Appendix – lattice rotations from mechanical constraint during micropillar compression

During a compression test, the pillar is mechanically constrained by the rigid flat-punch indenter face on top of the pillar, and by the rest of the sample at its base. The indenter constrains the pillar top face in the vertical direction, but the in-plane rotational and lateral constraint depends on friction between the pillar and indenter. The coefficient of friction is not known in this experiment.

Supplementary Figure 4 shows a similar basal pillar to the one in Figure 7, which rotated during compression. The pillar has bent so that the top part of the deformed pillar is laterally displaced from the base, but the top face has remained normal to the loading direction. This confirms that top of the pillar was vertically constrained and could not tilt out of plane, but laterally less constrained so that it could slide under the indenter. Since the top of the pillar can slide, but cannot tilt, crystal rotation from slip is likely to be minimum in this region.

EBSD analysis of a basal pillar deformed at 600° is described in part 2 of this study [40]. The crystal lattice near the top of the deformed pillar is within 5° of the undeformed orientation, which confirms the validity of this model.

# 9 Supplementary Figures

| Name | Slip dir / Burgers vec | Slip plane | TEM dislocations | Slip traces |
|---|---|---|---|---|
| ⟨c⟩ prism 1 | [0001] | {10$\bar{1}$0} | [17,25,29] | |
| ⟨c⟩ prism 2 | [0001] | {11$\bar{2}$0} | [25] | |
| ⟨a⟩ basal | $\frac{1}{3}$⟨11$\bar{2}$0⟩ | (0001) | [25] | |
| ⟨a⟩ prism1 | $\frac{1}{3}$⟨11$\bar{2}$0⟩ | {10$\bar{1}$0} | [17,25] | [17] |
| ⟨a⟩ pyr 1 | $\frac{1}{3}$⟨11$\bar{2}$0⟩ | {10$\bar{1}$1} | ([17]) | |
| ⟨$a_1 + a_2$⟩ pyr 2 | $\frac{1}{2}$⟨1$\bar{1}$00⟩ | {11$\bar{2}$2} | ([17]) | |
| ⟨c + a⟩ prism1 | $\frac{1}{3}$⟨11$\bar{2}$3⟩ | {10$\bar{1}$0} | [25,27,31] | [17,31] |
| ⟨c + a⟩ pyr 1 | $\frac{1}{3}$⟨11$\bar{2}$3⟩ | {10$\bar{1}$1} | ([17]) | |
| ⟨c + a⟩ pyr 2 | $\frac{1}{3}$⟨11$\bar{2}$3⟩ | {11$\bar{2}$2} | [32] ([17]) | |
| Prism 1 | unknown | {10$\bar{1}$0} | [17,25] | [17,24] |
| Prism 1 | [0001] component | {10$\bar{1}$0} | | [26,28,31] |
| Prism 1 | ⟨11$\bar{2}$0⟩ component | {10$\bar{1}$0} | | [26,28,31] |
| Pyr 2 | unknown | {11$\bar{2}$2} | [17] | |
| Pry 1 | unknown | {10$\bar{1}$1} | [17] | [17] |
| ⟨c + a⟩ partial | $\frac{1}{6}$⟨11$\bar{2}$3⟩ | {10$\bar{1}$0} Glissile stacking fault | [25–27,32] | - |
| ⟨$c + a_1 + a_2$⟩ partial | $\frac{1}{6}$⟨20$\bar{2}$3⟩ | {11$\bar{2}$2} stacking fault / antiphase boundary | [29,33] | - |
| ⟨a⟩ partial | $\frac{1}{6}$⟨11$\bar{2}$0⟩ | Sessile stair rod from reaction of ⟨c + a⟩ partials on adjacent {10$\bar{1}$0} planes | [17,27] | |
| ⟨c + a⟩ | $\frac{1}{3}$⟨11$\bar{2}$3⟩ | Screw/unknown | [17,25,32] | |
| ⟨c⟩ | [0001] | Screw/unknown | [25,32] | |
| ⟨a⟩ | $\frac{1}{3}$⟨11$\bar{2}$0⟩ | Screw/Unknown | [25,32,70] | |

*Supplementary Table 1: Summary of slip systems and dislocation types reported in WC at RT [17,24–29,31–33]. References in parentheses show where a slip system is reported but not directly observed from the experiment. Dislocation types and complete slip systems (i.e. both slip direction and slip plane) reported more than once are shown in red text.*

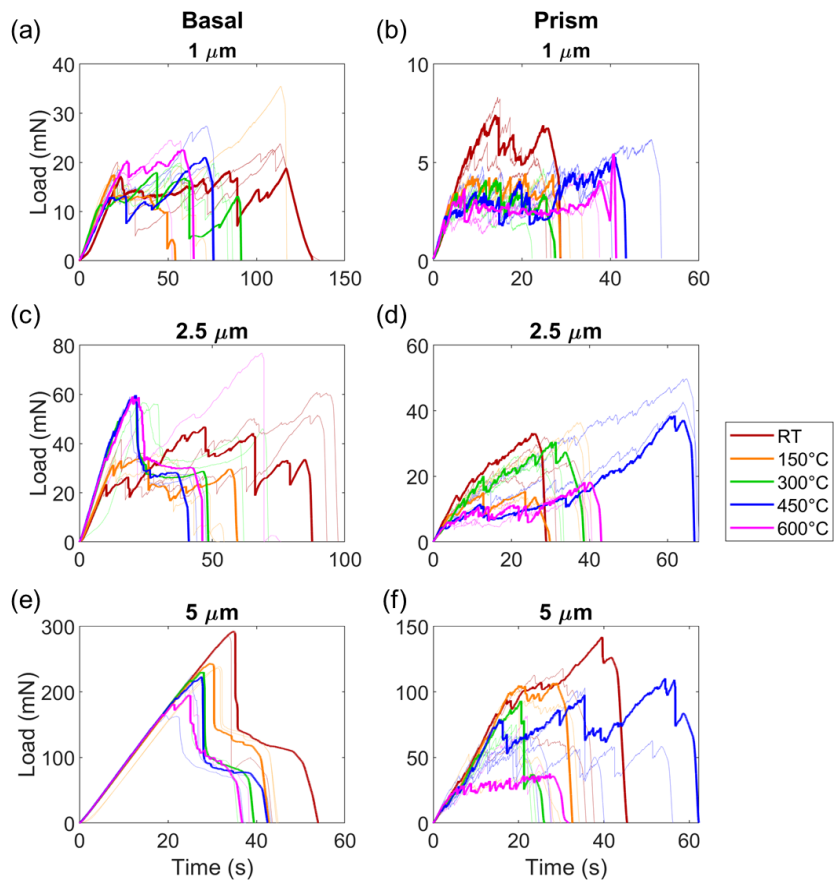

*Supplementary Figure 1: Load-time graphs for all pillars, grouped by pillar size (rows), loading orientation (columns), and test temperature (plot colours). One line per pillar type is plotted in thicker lines to help visualization.*

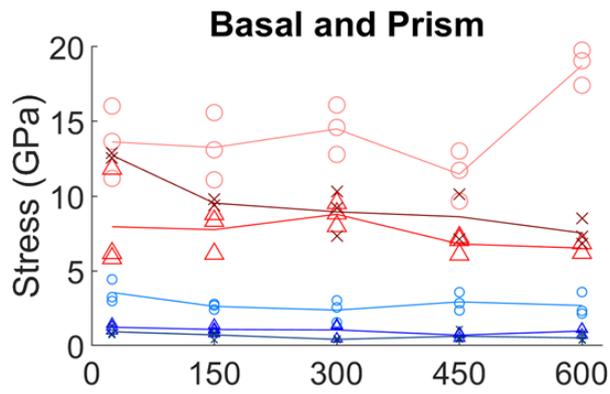
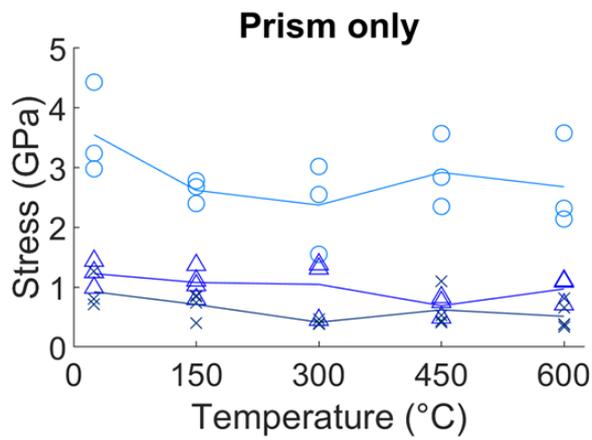
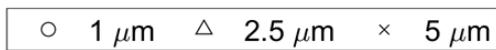

*Supplementary Figure 2: a) Variation of stress at first load drop with temperature, for each pillar orientation and size. Red lines/markers correspond to basal pillars, and blue lines/markers to prismatic pillars. Markers plot yield stress for individual pillars, and lines show the mean value for each pillar type. b) Prismatic oriented pillar data only from a) plotted on a re-scaled vertical axis.*

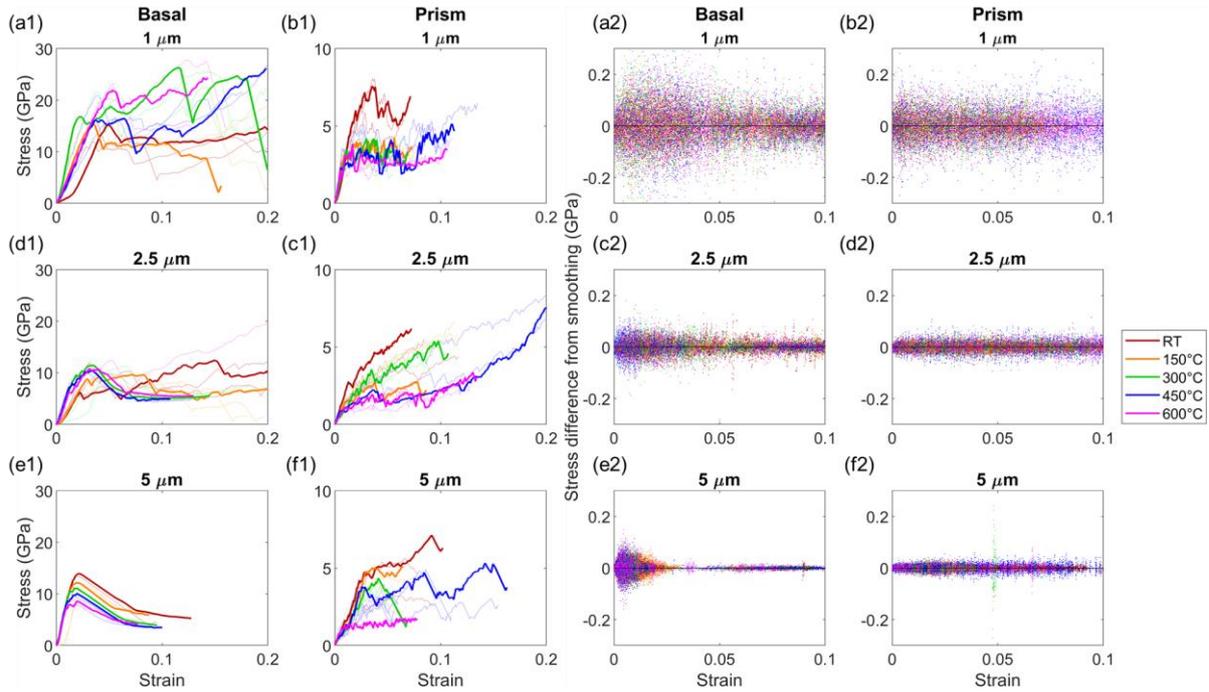

*Supplementary Figure 3: A Lowess smoothing filter was applied to the data to suppress small high-frequency fluctuations in stress and strain from the electronic noise of the support hardware before computing the stress-strain gradient. **(a1-f1)** shows the filtered stress-strain curves. **(a2-f2)** shows the difference between the smoothed and original data (plotted in Figure 2). The stress differences between the original and smoothed data are small (99.9% of points < 0.2 GPa) and decrease with increasing pillar size.*

*The Lowess filter uses locally weighted linear regression to fit the stress-strain data to a linear spline [61]. A moving window size spanning 1% of the nearest neighbour points was used for these data.*

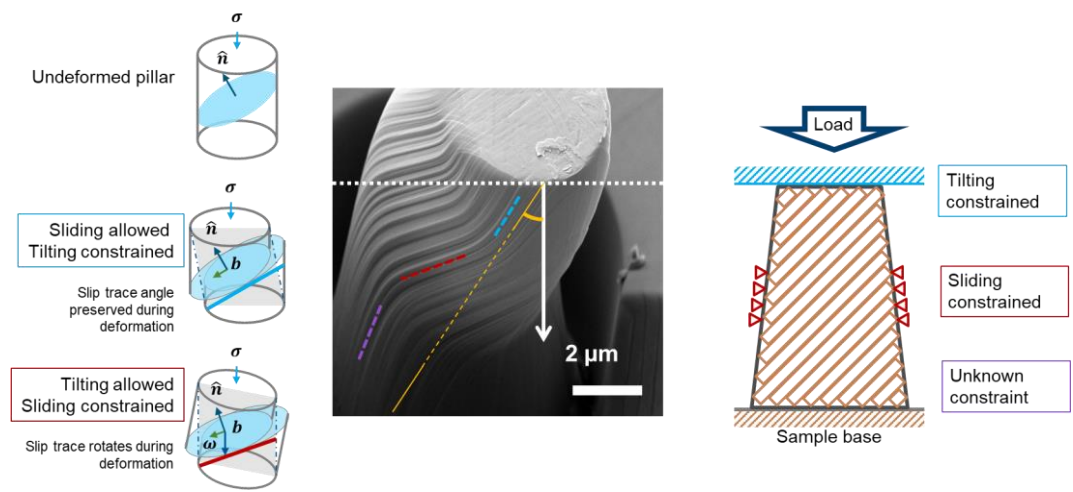

*Supplementary Figure 4: Mechanical constraint at the top and middle of a micropillar during compression. The top of the pillar cannot tilt but may slide laterally or rotate in-plane under the flat punch indenter tip. The middle of the pillar is constrained by the material around it so that it cannot slide, except where dislocations are emitted at the pillar surface.*

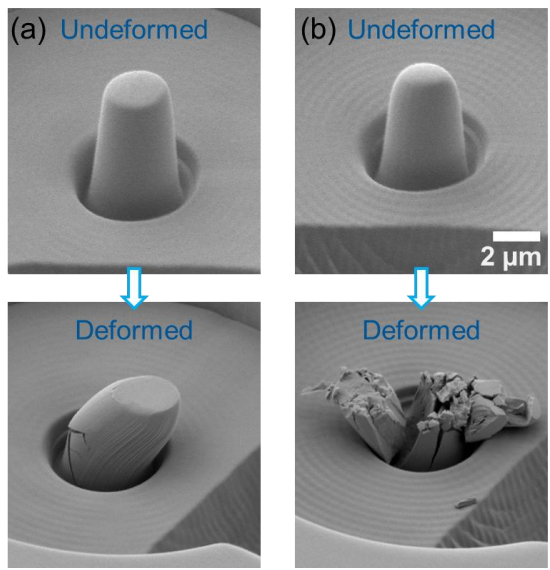

*Supplementary Figure 5: SEM images of 2.5 µm basal pillars before and after deformation at 600 °C. a) Pillar with less edge rounding deformed by deterministic stress-strain behaviour, unstable yielding and stable crack growth; b) Pillar with more edge rounding deformed by stochastic stress-strain behaviour and unstable axial cracks.*